\documentclass[twocolumn]{webofc}
\usepackage[varg]{txfonts}   % Web of Conferences font
\usepackage{lipsum,widetext}
\usepackage{multirow}
\usepackage{color}

\newcommand\beq{\begin{equation}}
\newcommand\eeq{\end{equation}}
\newcommand\beqa{\begin{eqnarray}}
\newcommand\eeqa{\end{eqnarray}}

\newcommand{\dd}{\text{d}}

\begin{document}
\title{Non-monotonic Mpemba effect in binary molecular suspensions}

\author{\firstname{Rub\'en} \lastname{G\'omez Gonz\'alez}\inst{1}\fnsep\thanks{\email{ruben@unex.es}} \and
        \firstname{Vicente} \lastname{Garz\'o}\inst{2}\fnsep\thanks{\email{vicenteg@unex.es}} 
}

\institute{Departamento de F\'isica, Universidad de Extremadura, E-06071 Badajoz, Spain 
\and
           Departamento de F\'isica and Instituto de Computaci\'on Cient\'ifica Avanzada (ICCAEx), Universidad de Extremadura, E-06071 Badajoz, Spain }

\abstract{%
  The Mpemba effect is a phenomenon in which an initially hotter sample cools sooner. In this paper, we show the emergence of a non-monotonic Mpemba-like effect in a molecular binary mixture immersed in a viscous gas. Namely, a crossover in the temperature evolution when at least one of the samples presents non-monotonic relaxation. The influence of the bath on the dynamics of the particles is modeled via a viscous drag force plus a stochastic Langevin-like term. Each component of the mixture interchanges energy with the bath depending on the mechanical properties of its particles. This discrimination causes the coupling between the time evolution of temperature with that of the partial temperatures of each component. The non-monotonic Mpemba effect---and its inverse and mixed counterparts---stems from this coupling. In order to obtain analytical results, the velocity distribution functions of each component are approximated by considering multitemperature Maxwellian distributions. The theoretical results derived from the Enskog kinetic theory show an excellent agreement with direct simulation Monte Carlo (DMSC) data.
}
\maketitle
\section{Introduction}
\label{intro}
Under certain conditions, a sample of water at a hotter temperature freezes faster. This  counterintuitive phenomenon is known in literature as the Mpemba effect according to E. B. Mpemba \cite{MO69}, who was the first researcher to make rigorous observations. Although several mechanisms have been proposed to explain this effect in water \cite{J06}, there are still concerns regarding the true nature of it \cite{BH20}. 

Leaving aside the origin of the Mpemba effect in water, it is clear that to observe this effect, the time evolution of temperature must be coupled to that of microscopic---or kinetic---variables. For this reason, the Mpemba effect is considered to be a memory effect \cite{KP19}. This kind of out-of-equilibrium processes have also been observed in various systems such as clathrate hydrates \cite{AK16}, polymers \cite{HL18}, and colloids \cite{KB20}. In these systems, a Mpemba-like effect emerges when two samples at different initial temperatures evolve in such a way that at a given time $t_c$ the relaxation curves for the temperatures cross each other. Thus, freezing is avoided and a straightforward explanation can be provided. 

However, there remain a lot of variables that have influence on the temperature evolution. As a consequence, setting the initial conditions giving rise to a crossing of the cooling curves can result in an arduous task. For this reason, from an analytical point of view, simple models have been usually considered to gain some insight into the problem. This is particularly the case of kinetic-theory-based models employed to describe the emergence of Mpemba-like effect (and its inverse counterpart, namely when the two initial temperatures are below the asymptotic one) in granular \cite{LR17,TC19,BP20,TH20}---i.e. with inelastic collisions--- and molecular gases \cite{SP20,GK20}. In the former case, inelasticity of collisions couples the evolution of the (granular) temperature with that of non-hydrodynamic variables--such as the fourth cumulant or \textit{kurtosis} and/or the rotational-to-translational temperature ratio. This coupling is the reason of the occurrence of the Mpemba effect in granular gases during the relaxation towards the steady state. On the other hand, no inelasticity couples the temperature with any kinetic variable in the case of molecular gases, therefore other mechanisms must be considered. For example, in a recent paper \cite{SP20}, the molecular gas is driven by means of a non-linear drag force plus a stochastic force. The non-linearity in the velocity dependence of the drag force turns out in the coupling between the temperature $T$---seen as a measure of the amount of kinetic energy---with the fourth cumulant $a_2$. Nevertheless, in concordance with the granular case \cite{G19}, $a_2$ is assumed to be small. As a result, initial temperatures must be chosen to be close enough to each other for the emergence of the Mpemba effect.

A stronger Mpemba-like effect has been recently reported in the case of a mixture of molecular gases \cite{GK20}. In this paper, the mixture is in contact with a thermal reservoir. As usual \cite{KH01}, in the context of the Enskog kinetic theory, the influence of the surrounding fluid on the solid particles is modeled via an external force composed by two terms: (i) a viscous drag force proportional to the (instantaneous) velocity of the solid particles and (ii) a stochastic Langevin-like term. While the first term tries to mimic the loss of energy due to frictional forces through the drag coefficients $\gamma_i$ ($i=1,2$), the second term simulates the energy transmission due to instantaneous and random collisions with the bath. In this work, in accordance with the results obtained in lattice-Boltzmann simulations \cite{YS09}, the drag coefficients $\gamma_i$ are chosen for each species. This differentiation in the interaction with the background gas causes the coupling between the evolution equations of the temperature $T(t)$ and the partial temperatures $T_i(t)$ of each component. This coupling allows that, with an appropriate choice of the initial values, the Mpemba effect (and its inverse equivalent) arises in the relaxation towards equilibrium---defined in terms of a Maxwellian distribution at the background temperature $T_\text{ex}$. Moreover, the use of the partial temperatures as the control parameter permits more flexibility in the selection of the initial conditions. Namely, temperature curves can cross even when the initial differences are of the same order than the temperature themselves (\textsl{large} Mpemba effect). Furthermore, a Maxwellian approximation was used to estimate the partial production rates $\xi_i$---which measure rate of energy interchanged in collisions among particles of species $i$ with $j$. Thus, no cumulants are needed and the effect is easily understood. 

The main goal of the present work is to extend the results obtained in \cite{GK20} to those situations where the temperature presents a non-monotonic relaxation through an appropiately choice of the initial conditions. The non-monotonic Mpemba-like effect appears when at least one of the two samples whose temperature curves cross presents non-monotonic relaxation. Namely, when the temperature difference $|T(t)-T_\text{ex}|$---of at least one of the samples---increases at the early stages of the evolution but, at a given time, $T(t)$ reaches the equilibrium value $T_\text{ex}$. To fulfil this goal, we consider the Enskog kinetic theory in combination with the Fokker--Planck suspension model described above. Since the theoretical predictions derived in this work are based on the use of the Maxwellian approach, a comparison between theory and simulations is convenient to assess the reliability of the theory. Here, we compare our theoretical results against the numerical solution of the kinetic equations by means of the DSMC method conveniently adapted to the suspension model \cite{B94,MG02}.

\section{Enskog kinetic theory}
\label{sec-1}
%For bibliography use \cite{RefJ}

Let us consider an ensemble of hard spheres of masses $m_1$ and $m_2$ and diameters $\sigma_1$ and $\sigma_2$ immersed in a viscous gas at temperature $T_\text{ex}$. At a kinetic-theory level, all the relevant information of the system is enclosed in the one-particle velocity distribution functions $f_i(\mathbf{r},\mathbf{v},t)$ of species $i$. For moderate densities and \emph{homogeneous} states, the functions $f_i$ obey the set of coupled Enskog kinetic equations \cite{CC70}:
\beq
\label{1}
\frac{\partial f_i}{\partial t}+\mathcal{F}_i f_i=\sum_{j=1}^2J_{ij}[\mathbf{r},\mathbf{v}|f_i,f_j], \quad i=1,2
\eeq    
where $J_{ij}[f_i,f_j]$ is the Enskog collision operator and the operator $\mathcal{F}_i$ represents the influence of the bath on the dynamics of the particles. For low-Reynolds numbers, this force term is usually represented by a drag force plus a Fokker--Planck collision operator \cite{NE98}. In this way, the Enskog equation reads \cite{GK20b}
\beq
\label{2}
\frac{\partial f_i}{\partial t}-\gamma_i\frac{\partial}{\partial\mathbf{v}}\cdot\mathbf{v}f_i-\frac{\gamma_ik_B T_{\text{ex}}}{m_i}\frac{\partial^2 f_i}{\partial v^2}=\sum_{j=1}^2\; J_{ij}[f_i,f_j],
\eeq
where $\gamma_i$ are the drag coefficients and $k_B$ is the Boltzmann constant. Here, according to the simulation results in gas-solid flows reported by Yin and Sundaresan \cite{YS09}, the coefficients $\gamma_i$ can be written as $\gamma_i=\gamma_0R_i$, where $\gamma_0=36\eta_g/\left[\rho(\sigma_1+\sigma_2)\right]$, $\rho=\sum_i m_i n_i$ is the total mass density, $n_i$ is the number density of the component $i$, and $\eta_g$ is the viscosity of the solvent. Moreover, dimensionless functions $R_i$ depend on the mole fraction $x_i=n_i/(n_1+n_2)$, mass $m_1/m_2$ and size $\sigma_1/\sigma_2$ ratios, and the partial volume fractions $\phi_i=(\pi/6) n_i\sigma_i^3$, related with the total one by $\phi=\phi_1+\phi_2$. More details of this kind of Langevin-like models can be found in Refs. \cite{HT17,GK20b}.

For homogeneous situations, all the relevant information of the binary mixture is provided by the partial temperatures $T_i(t)$ of the component $i$---or, equivalently, by the temperature ratio $\theta(t)=T_1(t)/T_2(t)$ and the (total) temperature $T(t)=x_1T_1(t)+x_2T_2(t)$ of the mixture. The partial temperatures are defined as
\beq
\label{4}
T_i=\frac{1}{3 k_B n_i}\int \dd \mathbf{v}\; m_i v^2\; f_i(\mathbf{v}).
\eeq

The evolution equations for the temperature ratio $\theta$ and the (reduced) temperature $T^*=T/T_\text{ex}$ can be obtained by multiplying both sides of the Enskog equation \eqref{2} by $m_i v^2$ and integrating over velocity. They are given by \cite{GK20}\vspace{4mm}
\begin{widetext}
\beq
\label{5}
\frac{\partial}{\partial t^*} T^*=2\left(x_1\gamma_1^*+x_2 \gamma_2^*\right)-2T^* \frac{x_1 \gamma_1^* \theta+x_2 \gamma_2^*}{1+x_1(\theta-1)}\equiv\Phi,
\eeq
\beq
\label{6}
 \frac{\partial}{\partial t^*} \theta=2\left(\gamma^*_1-\gamma_2^*\theta\right)\frac{1+x_1(\theta-1)}{ T^*}-2\theta(\gamma_1^*-\gamma_2^*)+\frac{16\sqrt{\pi}}{3}\chi_{12}\sqrt{\frac{T^*}{2}\frac{\mu_{12}\mu_{21}\left(\mu_{12}+\mu_{21}\theta\right)}{1+x_1(\theta-1)}}
 \left(1-\theta\right)\left(x_1\theta+x_2\right),
\eeq
where we have introduced the scaled time $t^*=4(n_1+n_2)/(\sigma_1+\sigma_2)^2\sqrt{4k_BT_\text{ex}(m_1+m_2)}t$. Here, $\mu_{ij}=m_i/(m_i+m_j)$, $\chi_{12}$ is the pair correlation function,
\beq
\label{7}
\gamma_i^*=\frac{8 R_i}{\sqrt{2T_\text{ex}^*}(n_1+n_2)(\sigma_1+\sigma_2)^3}, \quad \text{and}\quad T_\text{ex}^*=\frac{8k_BT_\text{ex}}{(m_1+m_2)(\sigma_1+\sigma_2)^2\gamma_0^2}.  
\eeq
\end{widetext}
In order to write Eq~\ref{6}, the partial production rates 
\beq
\label{8}
\xi_i=-\frac{m_i}{3 n_i k_B T_i}\int \dd \mathbf{v}\; v^2 J_{ij}[f_i,f_j], \quad i \neq j
\eeq
 have been calculated by approaching the distribution functions $f_i$ to their Maxwellian forms:
\beq
\label{9}
f_i(\mathbf{v},t)\to n_i \left(\frac{m_i}{2\pi k_B T_i(t)}\right)^{3/2} \exp\left(-\frac{m_i v^2}{2k_BT_i(t)}\right). 
\eeq
More details can be found in Refs. \cite{G67,GK20}.

If we freely let the mixture evolve, it will asymptotically achieve the equilibrium state where equipartition holds, i.e.  $T^\text{eq}_{1}=T^\text{eq}_{2}=T_\text{eq}=T_{\text{ex}}$. However, during the time evolution towards this final state, partial temperatures are not equal ($T_1(t)\neq T_2(t)$). This symmetry breaking and the fact that $\gamma_1\neq\gamma_2$ (because $R_1\neq R_2$) causes the emergence of the Mpemba effect in the mixture. Note that in the case that $\gamma_1=\gamma_2$, total temperature evolution decouples from that of the partial temperatures and the Mpemba effect dissapears.

\section{Results and discussion}
\label{sec-2}
In this section, we consider temperature crossings where a non-monotonic relaxation is present. For this reason, the non-linearity of the evolution equations \ref{5}-\ref{6} plays a crucial role. The linear counterpart has been previously analysed in Ref. \cite{GK20} where a perfect agreement between $\emph{exact}$ theoretical expressions---for the (scaled) crossing time $t^*_c$---and DSMC simulations has been found. Unfortunately, unlike the linear case, we are lack of simple analytical expressions in the far-from-equilibrium framework. Thus, more qualitative analysis is required to establish the necessary but no sufficient conditions for the ocurrence of the Mpemba effect. 

Let us consider two homogeneous states A and B far away from equilibrium. The time evolution of these states is fully determined by their initial temperatures $T^*_{0,\text{A}}$ and $T^*_{0,\text{B}}$ and their temperature ratios $\theta_{0,\text{A}}$ and  $\theta_{0,\text{B}}$. Thus, we can establish a necessary condition---for a crossover in the temperature relaxation---based on the selection of the initial slopes $\Phi(T^*_{0,\text{A}},\theta_{0,\text{A}})$ and $\Phi(T^*_{0,\text{B}},\theta_{0,\text{B}})$ of each state through their dependence on the initial values $T^*_0$ and $\theta_0$. Under this assumption, a necessary condition for the Mpemba effect is $\Phi(T^*_{0,\text{B}},\theta_{0,\text{B}})>\Phi(T^*_{0,\text{A}},\theta_{0,\text{A}})$, where we have considered the state A to be initially hotter than B, namely $T^*_{0,\text{A}}>T^*_{0,\text{B}}$.

Next step is to ensure that the function $\Phi(T^*,\theta)$ exhibit a monotonic dependence on $\theta$, so that, with an appropriate selection of the temperature ratios $\theta_{0,\text{A}}$ and $\theta_{0,\text{B}}$, we can ensure the temperatures $T^*_\text{A}$ and $T^*_\text{B}$ will get closer or away from each other at the early stages. Only the first option is considered here as a simple way to attain the Mpemba effect. This will be always the case for those systems that behave monotonically and for those with non-monotonic relaxation but in which temperature evolution presents only one relative extreme value. The aim of this paper is the analysis of the latter situation. However, the determination of some criterion on whether or not temperature evolution behaves monotonically with the temperature ratio falls outside the purposes of this work.

Hence, to test the emergence of the Mpemba effect we perform the derivative of $\Phi$ with respect to $\theta$ for fixed $T^*$. The result is
\beq
\label{10}
\left(\frac{\partial \Phi}{\partial\theta}\right)_{T^*}=2T^*\frac{x_1x_2(\gamma_2^*-\gamma_1^*)}{(x_2+x_1\theta)^2},
\eeq
which is always a positive (negative) function if $\gamma_2^*>\gamma_1^*$ ($\gamma_2^*<\gamma_1^*$). Therefore, the initial values must satisfy the following conditions
\beqa
\label{11}
\frac{T^*_{0,\text{A}}-T^*_{0,\text{B}}}{\theta_{0,\text{A}}-\theta_{0,\text{B}}}&>&0, \quad  \gamma_1^*>\gamma^*_2,\nonumber\\
\frac{T^*_{0,\text{A}}-T^*_{0,\text{B}}}{\theta_{0,\text{A}}-\theta_{0,\text{B}}}&<&0, \quad  \gamma_1^*<\gamma^*_2.
\eeqa
Same conditions were obtained in the linear analysis developed in Ref. \cite{GK20}. However, Eq.~\ref{11} does not constrain the region the initial conditions must belong to. In other words, differences between $\Phi(T^*_{0,\text{A}},\theta_{0,\text{A}})$ and $\Phi(T^*_{0,\text{B}},\theta_{0,\text{B}})$ must be chosen to be large enough for the occurrence of the Mpemba effect.
\begin{figure}[h!]
	\centering
	\includegraphics[width=0.45\textwidth]{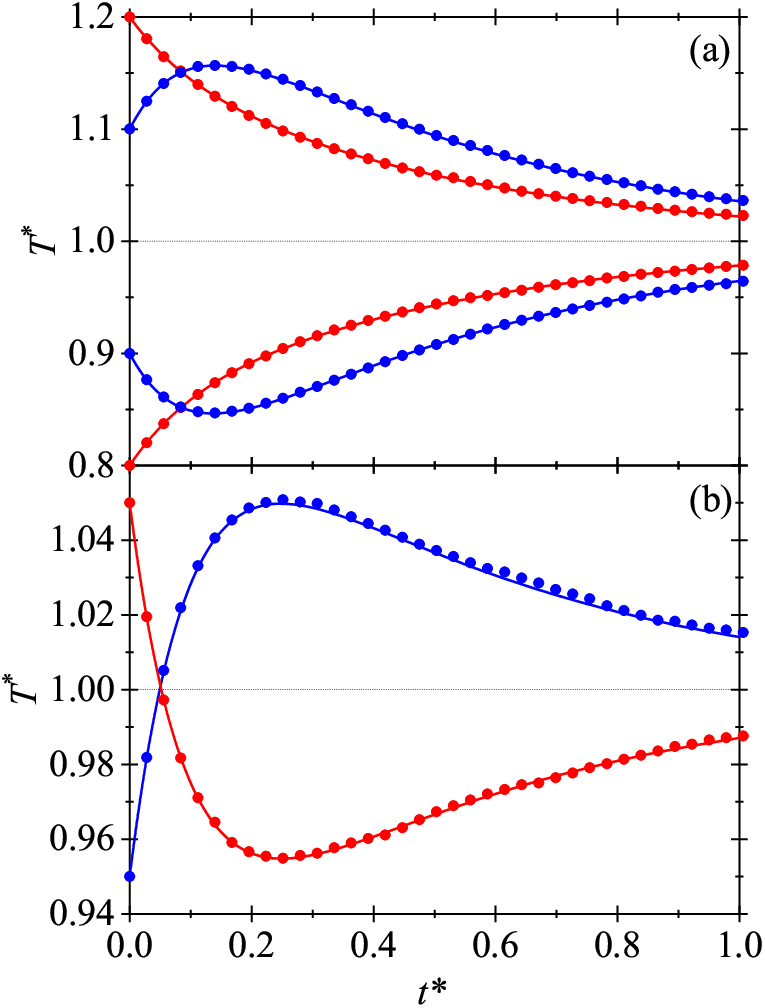}
	%\vspace*{5cm}
	\caption{Evolution of the (reduced) temperature $T^*$ over the time $t^*$ for  $m_1/m_2=10$, $\sigma_1/\sigma_2=1$, $\phi=0.1$, $T_\text{ex}^*=1$, and $x_1=0.5$. Solid lines represent theoretical values and symbols DSMC data.  From  top  to bottom, panel (a) corresponds to the NMME (cooling) and to the NMIME (heating) and panel (b) to the MME.}
	\label{fig-1}       % Give a unique label
\end{figure}

\begin{table}[h]
\centering
\caption{Initial values of the (reduced) temperatures $T^*_0$ and temperature ratios $\theta_0$ used to generate the relaxation curves shown in Fig.~\ref{fig-1}.}
	\begin{tabular}{|c|c|c|c|c|}
		\hline
		& \multicolumn{2}{c|}{Panel (a)} & \multicolumn{2}{c|}{Panel (b)} \\ \hline
		Color of lines and symbols & $T^*_0$      & $\theta_0$      & $T^*_0$      & $\theta_0$      \\ \hline
		& \multicolumn{2}{|c|}{NMME}    & \multicolumn{2}{c|}{MME}                                                      \\ \cline{1-5}
		Red                        & 1.2          & 1.1             & 1.05          & 0.6             \\ \cline{1-5}
		Blue                       & 1.1          & 2.1             & \multicolumn{2}{c|}{\multirow{2}{*}{}}           \\ \cline{1-3}
		&\multicolumn{2}{|c|}{NMIME}          	&\multicolumn{2}{c|}{---}                                                     \\ \cline{1-3}
		Red                        & 0.8          & 0.9            & \multicolumn{2}{c|}{}            \\ \cline{1-5}
		Blue                       & 0.9          & 0.4             & 0.95          & 1.8             \\ \cline{1-5}
	\end{tabular}
	\label{tab-1}
\end{table}

To illustrate the occurrence of the non-monotonic Mpemba effect, we consider the following approach for the pair correlation functions \cite{B70}
\beq
\label{20.1}
\chi_{ij}=\frac{1}{1-\phi}+\frac{3\phi}{(1-\phi)^2}\frac{\sigma_i\sigma_jM_2}{(\sigma_i+\sigma_j)M_3}+
\frac{2\phi^2}{(1-\phi)^3}\left(\frac{\sigma_i\sigma_jM_2}{(\sigma_i+\sigma_j)M_3}\right)^2,
\eeq
where $M_\ell=\sum_i x_i\sigma_i^\ell$.

The non-monotonic Mpemba effect for a cooling, a heating and a \textit{mixed} transition are plotted in Fig.~\ref{fig-1}. More specifically, the mixture under consideration is composed of equally distributed ($x_1=\frac{1}{2}$) hard spheres of same diameters ($\sigma_1=\sigma_2$) and different masses ($10m_1=m_2$) . In addition, we consider a moderate dense system ($\phi=0.1$). Lines are the theoretical results derived from the Enskog equation and symbols represent the DSMC data. For this parameter space, $\gamma_1^*=0.445$ and $\gamma_2^*=4.451$ and, consequently, the ratio $\left(T^*_{0,\text{A}}-T^*_{0,\text{B}}\right)/\left(\theta_{0,\text{A}}-\theta_{0,\text{B}}\right)$ is chosen to be less than 0 in concordance with Eq.~\ref{11}. For the Mpemba effect to occur, slopes $\Phi(T^*,\theta)$ are independently selected for the cooling and the heating cases (see table~\ref{tab-1} for more details). Fig.~\ref{fig-1}(a) shows the non-monotonic Mpemba effect (NMME)---and its inverse counterpart (NMIME)---to emerge even when the initial temperature differences are around 10\% (\emph{large} non-monotonic Mpemba effect). On the other hand, in Fig.~\ref{fig-1}(b) the system is initiated at two different temperatures: one cooler and the other hotter than $T_\text{ex}$. Because of the non-linearity of the evolution equation of the temperature, a crossover is still possible. This curious effect is the so-called mixed Mpemba effect (MME) \cite{TH20}. Moreover, Fig.~\ref{fig-1} highlights an excellent agreement between the Enskog theory and the DSMC simulations. This excellent agreement ensures the accuracy of the Maxwellian approximation (Eq.~\ref{9}) to model the velocity distribution function in these out-of-equilibrium systems.
%
%\section*{Acknowledgments}

\small The  present  work  has  been  supported  by the Spanish government through Grant No. FIS2016-76359-P and by the Junta de Extremadura (Spain) through Grant Nos. IB16013 and GR18079, partially financed by “Fondo Europeo de Desarrollo Regional” funds. The research of Rub\'en G\'omez Gonz\'alez was supported  by  the  predoctoral  fellowship  of  the  Spanish  government, Grant No. BES-2017-079725.
\\


\begin{thebibliography}{24}
\bibitem{MO69}
E.B. Mpemba, D.G. Osborne, Phys. Educ. \textbf{4}, 172-175 (1969)

\bibitem{J06}
M. Jeng, AM. J. Phys. \textbf{74}, 514-522 (2006)

\bibitem{BH20}
H.C. Burridge, O. Hallstadius, Proc. Royal. Soc. A \textbf{476}, 20190829 (2020)

\bibitem{KP19}
N.C. Keim, J.D. Paulsen, Z. Zeravcic, S. Sastry, S.R. Nagel, Rev. Mod. Phys. \textbf{91}, 035002 (2019)

\bibitem{AK16}
Y. Ahn, H. Kang, D. Koh, H. Lee, Korean J. Chem. Eng. \textbf{33}, 1903-1907 (2016)

\bibitem{HL18}
C. Hu, J. Li, S. Huang, H. Li, C. Luo, J. Chen, S. Jiang, L. An, Cryst. Growth Des. \textbf{18}, 5757-5762 (2018)

\bibitem{KB20}
A. Kumar, J. Bechhoefer, Nature \textbf{584}, 64-68 (2020)

\bibitem{LR17}
A. Lasanta, F.V. Reyes, A. Prados, A. Santos, Phys. Rev. Lett. \textbf{119}, 148001 (2017)

\bibitem{TC19}
A. Torrente, M.A. L\'opez-Casta\~no, A. Lasanta, F.V. Reyes, A. Prados, A. Santos, Phys. Rev. E \textbf{99}, 060901(R) (2019)

\bibitem{BP20}
A. Biswas, V.V. Prasad, O. Raz, R. Rajesh, Phys. Rev. E \textbf{102}, 012906 (2020)

\bibitem{TH20}
S. Takada, H. Hayakawa, A. Santos, arXiv: 2011.00812 (2020)

\bibitem{SP20}
A. Santos, A. Prados, Phys. Fluids \textbf{32}, 072010 (2020)

\bibitem{GK20}
R. G\'omez Gonz\'alez, N. Khalil, V. Garz\'o, Phys. Fluids \textbf{33}, 053301 (2021)

\bibitem{G19}
V. Garz\'o, \textit{Granular Gaseous Flows} (Springer Nature Switzerland, Basel, 2019)


\bibitem{KH01}
 D.L. Koch, R.J. Hill, Annu. Rev. Fluid Mech. \textbf{33}, 619-647 (2001)
 
\bibitem{YS09}
X. Yin, S. Sundaresan, AIChE \textbf{55}, 1352-1368 (2009)

\bibitem{B94}
G.A. Bird, \textit{Molecular Gas Dynamics and the Direct ~Simulation of Gas Flows} (Oxford University Press, Oxford, 1994)

\bibitem{MG02}
J.M. Montanero, V. Garz\'o, Granul. Matter \textbf{4}, 17-24 (2002)

\bibitem{CC70}{S. Chapman, T.G. Cowling, \textit{The Mathematical Theory of Non-Uniform Gases} (Cambridge University Press, Cambridge, 1970)}

\bibitem{NE98}
T.P.C. van Noije, M.H. Ernst, Granul. Matter \textbf{1}, 57-64 (1998)

\bibitem{GK20b}
R. G\'omez Gonz\'alez, N. Khalil, V. Garz\'o, Phys. Rev. E \textbf{101}, 012904 (2020)

\bibitem{HT17}
H. Hayakawa, S. Takada, V. Garz\'o, Phys. Rev. E \textbf{96},  069904 (2017)

\bibitem{G67}
E. Goldman, Phys. Fluids \textbf{10}, 1928 (1967)

\bibitem{B70}
T. Boubl\'ik, J. Chem. Phys. \textbf{53}, 471-472 (1970)
\end{thebibliography}
\end{document}